\documentclass[12pt]{scrartcl} 
\usepackage{a4}
\usepackage{empheq}
\usepackage{amsthm}
\usepackage{amssymb}
\usepackage{amsfonts}
\usepackage{mathrsfs}
\usepackage{latexsym}
\usepackage{color}
\usepackage{bbm}
\usepackage[a4paper,colorlinks,citecolor=blue,linkcolor=blue,urlcolor=blue,pdfpagemode=None]{hyperref}
\usepackage[square,numbers,sort&compress]{natbib}
\usepackage{alltt}
\usepackage{framed}
\usepackage{multirow}
\usepackage{mathdots}

  \vfuzz \hfuzz
\newcommand{\D}{\text{d}}

\newcommand{\I}{\text{i}}
\newcommand{\C}{\mathbbm{C}}

\newcommand{\Z}{\mathbbm{Z}}
\def\1{\ifmmode\mathrm{1\!l}\else\mbox{\(\mathrm{1\!l}\)}\fi}
\newcommand{\one}{\mathbbm{1}}
\newcommand{\be}{\begin{equation}}
\newcommand{\ee}{\end{equation}}
\newcommand{\bes}{\begin{equation*}}
\newcommand{\ees}{\end{equation*}}

\newcommand{\Ainf}{A_\infty}

\allowdisplaybreaks

\deffootnote[1em]{1em}{1em}
{\textsuperscript{\thefootnotemark}}

\theoremstyle{definition}
\newtheorem*{definition}{Definition}
\newtheorem*{proposition}{Proposition}
\newtheorem*{theorem}{Theorem}

\numberwithin{equation}{section}

\begin{document}

\title{Matrix factorisations and open topological string theory}
\author{Nils Carqueville
\\[0.5cm]
  {\normalsize\slshape King's College London, Department of Mathematics,}\\[-0.1cm]
  {\normalsize\slshape Strand, London WC2R\,2LS, UK}\\[-0.1cm]
  \normalsize{\tt \href{mailto:nils.carqueville@kcl.ac.uk}{nils.carqueville@kcl.ac.uk}}}
\date{}
\maketitle

\begin{abstract}
Amplitudes in open topological string theory may be described completely by certain $\Ainf$-categories. We detail a general construction of all cyclic minimal models for a given $\Ainf$-algebra and apply this result to the case of $\mathcal N=2$ supersymmetric Landau-Ginzburg models. This allows to solve the tree-level theory in the sense that all amplitudes and hence the effective superpotential can be computed algorithmically. Furthermore, the construction provides a novel derivation of the topological metric of such models.
\end{abstract}

\section{Introduction}\label{introduction}

There are several reasons to study the underlying structure of topological string theory. On the one hand, it provides a rather versatile tool to compute certain quantities from full string theory (to whose chiral sector the topological theory is equivalent). Furthermore, though being far from trivial, topological string theory provides a more clear-cut arena to identify fundamental relations, explicify intuitions, and to structure them in a precise and rigorous language. Such a well-defined setting is important for at least three reasons: it enhances computational prowess in concrete problems; it serves as a solid and reliable platform from which to generalise to full string theory and try to arrive at new results and insights there; and finally, topological string theory forms a valuable  bilaterally permeable junction to pure mathematics, both making new techniques available to physics and providing mathematics with unexpected structures and relations to investigate and enjoy. 

This is true in particular for \textit{closed} topological string theory and its relation to enumerative and algebraic geometry, e.\,g.~counting instantons by Gromov-Witten invariants. In more recent years, also much progress has been made in the \textit{open} sector, describing D-branes and their spectra. Their topological properties are encoded in D-brane categories~\cite{d0011,s9902,al0104,l0305095} whose objects and morphisms describe branes and open strings, respectively. Such categories are endowed with very useful and very deep structure which is also at the heart of the homological mirror symmetry programme~\cite{k9411018}.

More specifically, in the present paper we will be concerned with the boundary sector of the topological B-twist~\cite{eguchiyang1990,w9112,v1991} of certain $\mathcal N=2$ superconformal theories that describe the stringy regime in K\"{a}hler moduli space for a wide range of type IIB compactifications. The ring of chiral primary fields on the boundary is equivalent to the BRST cohomology with basis~$\{\psi_i\}$ of the twisted theory. As a consequence, amplitudes $Q_{i_1\ldots i_n}$ with integrated descendants $\int\psi_i^{(1)}=\int[G_{-1}\D z+\bar G_{-1}\D\bar z,\psi_i]$ that are computed in the topological sector, 
\be\label{amplitudes}
Q_{i_1\ldots i_n} = (-1)^{|\psi_{i_1}|+\ldots+|\psi_{i_n}|+n} \Big\langle \psi_{i_1}\psi_{i_2}\mathcal P\int\psi_{i_3}^{(1)}\ldots \int\psi_{i_{n-1}}^{(1)} \, \psi_{i_n} \Big\rangle_{\text{disk}} \, ,
\ee
allow~\cite{bdlr9906200} to compute effective F-term superpotentials of the full string theory:
$$
\mathcal W_{\text{eff}} = \sum_{n\geq 2}\frac{1}{n+1}Q_{i_0 i_1\ldots i_n}u_{i_0}u_{i_1}\ldots u_{i_n}
$$
where the parameters $u_i$ have the opposite Grassmann parity to the fields $\psi_i$, and where $\langle\psi_i\psi_j\rangle_{\text{disk}}$ is the 2-point-correlator of open topological field theory on the disk, also known as the topological metric.

It was shown in~\cite{hll0402,hm0006120} that BRST symmetry implies that the amplitudes~(\ref{amplitudes}) coming from any $\mathcal N=2$ topological conformal field theory obey a family of constraint conditions that can be viewed as a pendant of the bulk WDVV equations in the boundary sector. Mathematically these constraints endow the open string spaces in the category of D-branes with the structure of $\Ainf$-algebras. Moreover, it follows from the Ward identities that the amplitudes~(\ref{amplitudes}) have a cyclic symmetry, and that 2-point-correlators $\langle\psi_i\psi_j\rangle$ are constant under deformations. As twisted topological conformal field theories are always cohomological field theories, this means that any $\mathcal N=2$ open topological string theory is naturally endowed with the structure of a cyclic, unital and minimal $\Ainf$-category. (See the next section for the precise definitions.)

Starting from a Segal-type definition of a topological conformal field theory the same structure was also obtained, and indeed shown to be equivalent with the definition itself, in~\cite{c0412149,LazaAinf}. In this sense we may identify open topological string theory with cyclic, unital and minimal $\Ainf$-categories, and this identification is what we shall exploit in the present paper. Emphasising the abstract algebraic $\Ainf$-structure of open topological string theory is not unlike allowing the most general characterisation of string vacua: the latter may be any abstract conformal field theory, irrespective of whether it has a known Lagrange formulation or a direct spacetime interpretation. 
This opens the door to all the techniques and insights of conformal field theory. Similarly, by stressing the general $\Ainf$-structure of open topological string theory one gains a deeper conceptual understanding. But on the other hand, we will see that the correct $\Ainf$-structure can be constructed explicitly, and this goes hand in hand with a computation of amplitudes and effective superpotentials ``from first principles'', i.\,e.~from the defining $\Ainf$-structure. 

The prototype example of a D-brane category is the one that describes branes in non-linear sigma models with a compact Calabi-Yau variety $X$ as its target space. It is by now rather well understood~\cite{s9902,d0011,al0104,hhp0803.2045} that the B-type boundary sector of such models is given by the bounded derived category $D(X)$ of coherent sheaves on $X$. As befits a cohomological field theory, this category is the cohomology category of a DG category $P(X)$.\footnote{If $\mathcal K(X)$ denotes the standard DG category of complexes of coherent sheaves on $X$, an object $P$ in $\mathcal K(X)$ is called h-projective iff $\text{Hom}_{H^0(\mathcal K(X))}(P,A)=0$ for all acyclic $A$ in $\mathcal K(X)$. One choice for $P(X)$ is then the full subcategory of all h-projectives in $\mathcal K(X)$.} Thus one may construct an $\Ainf$-structure on $D(X)$ induced from $P(X)$ as recalled in the next section, to wit, $D(X)$ is then a ``minimal model'' of $P(X)$. Furthermore such a construction can be guaranteed to be cyclic with respect to the topological metric. The reason is that the latter is induced from the pairing
\be\label{DXtopmetric}
\langle\alpha,\beta\rangle_\sigma = \int_X\Omega\wedge\text{tr}(\alpha\beta)
\ee
where $\Omega$ is a holomorphic top form and $\alpha,\beta$ are observables in the large volume limit. Hence $\langle\,\cdot\,,\,\cdot\,\rangle_\sigma$ is cyclic already off-shell, and this property will be inherited to cohomology. As a consequence, higher $\Ainf$-products and effective superpotentials may in principle be computed rather straighforwardly for such non-linear sigma models~\cite{ak0412209}. 

\vspace{0.2cm}

Another interesting class of theories are $\mathcal N=2$ supersymmetric Landau-Ginzburg models. It has been argued in~\cite{kms1989,vw1989,m1989,howewest1,howewest2,howewest3} how to obtain conformal field theories as their infra-red limit in the renormalisation group flow. This is particularly important when applied to Gepner models, which may then be alternatively described by certain Landau-Ginzburg theories. Also, such theories are equivalent to non-linear sigma models on hypersurfaces in projective space both in the bulk~\cite{w9301} and boundary~\cite{o0503,hhp0803.2045} sector. 

The boundary sectors of B-twisted Landau-Ginzburg models with potential $W$ (and with flat target spaces) are given by the D-brane category $\text{MF}(W)$ of matrix factorisations of $W$~\cite{kl0210,bhls0305,l0312}. If the Landau-Ginzburg model has $N$ chiral superfields $x_1,\ldots,x_N$, the objects of $\text{MF}(W)$ are pairs of square matrices $(d_0,d_1)$ with entries in $\C[X]:=\C[x_1,\ldots,x_N]$ that factorise the potential $W$. By combining $d_0$ and $d_1$ into $D:=(\begin{smallmatrix}0&d_1\\d_0&0\end{smallmatrix})$, this condition precisely means $D^2=W\cdot\one$. Given two such matrix factorisations $D$ and $D'$ of size $2r$ and $2r'$, respectively, we consider the space $V_{2r,2r'}$ of polynomial $(2r\times 2r')$-matrices. This space is naturally $\Z_2$-graded where block-diagonal matrices have degree 0 and off-block-diagonal matrices have degree 1. It follows from the matrix factorisation condition that the map $d_{DD'}\in\text{End}(V_{2r,2r'})$ defined on homogeneous $\phi$ by
$$
d_{DD'}(\phi) = D'\phi - (-1)^{|\phi|}\phi D
$$
is a differential, and one finds that it is the BRST operator on the boundary. Thus the morphism spaces $\text{Hom}(D,D')$ of $\text{MF}(W)$ are given by the cohomology $H_{d_{DD'}}(V_{2r,2r'})$, modelling open string states between the branes described by $D$ and $D'$. 

It is clear from this definition that $\text{MF}(W)$ is the cohomology category of the off-shell DG category $\text{DG}(W)$. The objects of the latter are also matrix factorisations, but $\text{Hom}_{\text{DG}(W)}(D,D')$ is equal to the full space $V_{2r,2r'}$ before taking cohomology with respect to the differential $d_{DD'}$. 

In order to establish that matrix factorisations of Landau-Ginzburg models are not only examples of mere open topological field theories given by $\text{MF}(W)$, but that they also have the full structure of open topological string theory, it is natural to try to obtain the proper $\Ainf$-structure on $\text{MF}(W)$ from the one on $\text{DG}(W)$. However, the naive construction of a generic minimal model quickly faces a serious problem. This arises because the correct $\Ainf$-structure on $\text{MF}(W)$ must be cyclic with respect to the topological metric which in the case of Landau-Ginzburg models does not have nice properties off-shell. Explicitly it was obtained in~\cite{kl0305,hl0404} by a boundary generalisation of the path integral derivation of~\cite{v1991} as the residue~\cite{GandH}
\be\label{KL}
\langle\phi_1,\phi_2\rangle_{\text{LG}}^D = \frac{1}{(2\pi\I)^N}\oint\frac{\text{str}(\partial_1 D\ldots\partial_N D\,\phi_1\phi_2)}{\partial_1 W\ldots\partial_N W}\D x_1\wedge\ldots\wedge\D x_N
\ee
for a brane $D$. One may easily check that on cohomology~(\ref{KL}) is well-defined and cyclic with respect to ordinary mutliplication, i.\,e.~$\langle\phi_1,\phi_2\phi_3\rangle_{\text{LG}}=\pm\langle\phi_2,\phi_3\phi_1\rangle_{\text{LG}}$, thus completing the structure of open topological field theory on $\text{MF}(W)$.\footnote{In all examples considered one finds that $\langle\,\cdot\,,\,\cdot\,\rangle_{\text{LG}}$ is non-degenerate, but a general proof of non-degeneracy has not been published.} However, the pairing~(\ref{KL}) is not cyclic off-shell, i.\,e.~in the category $\text{DG}(W)$. As will be discussed in more detail in the next section, this makes it much more difficult to construct an $\Ainf$-structure on $\text{MF}(W)$ whose higher products are cyclic with respect to the topological metric. But only cyclic products would allow to compute the amplitudes~(\ref{amplitudes}) or effective superpotentials, and only with cyclic products can matrix factorisations be endowed with the full structure of open topological string theory.\footnote{To ensure that the $\Ainf$-structure is unital does not turn out to be a problem.} 

\vspace{0.2cm}

The failure of~(\ref{KL}) to be a ``good'' pairing off-shell calls for a more sophisticated method to find the correct minimal $\Ainf$-structure on $\text{MF}(W)$. The construction that will be employed in the present paper crucially involves a reformulation of $\Ainf$-theory in terms of non-commutative geometry in the sense of Kontsevich~\cite{kontsevich1993,hl0410621,ks0606241} that was first used for topological string theory in~\cite{l0507222}. This approach effectively condenses all amplitudes and the topological metric into a single differential and a non-commutative symplectic form, thereby organising the complicated $\Ainf$-conditions in a clever way. This allows to discern very clearly in which precise way a generic (but easily constructible) minimal model fails to be cyclic and then ``correct'' it. As we will see in the body of the paper, this construction is entirely explicit and can be completely automatised. 

But not only can the treatment via non-commutative geometry provide cyclic $\Ainf$-products for a given pairing: without any additional effort, this approach allows to find and construct \textit{all} pairings with respect to which cyclic minimal models exist. In the case of Landau-Ginzburg models this means that one does not have to assume the topological metric~(\ref{KL}) but one can rather \textit{recover} it as one of the possible cyclic pairings. The fact that symplectic forms on a fixed space are all the same up to a choice of basis makes cyclic pairings essentially unique. This may be viewed as an alternative and path integral free derivation of the topological metric~(\ref{KL}) from first principles. 

\vspace{0.2cm}

In the present paper, we explain in detail the general construction of all cyclic minimal models for an arbitrary $\Ainf$-algebra, apply this method to Landau-Ginzburg models, and illustrate the recovery of the topological metric and the computation of effective superpotentials with a number of examples. 

In section~\ref{mathback}, we start by recalling the basics of $\Ainf$-algebras,\footnote{Any sufficiently small $\Ainf$-category $\mathcal A$ can be ``summed up'' to give an $\Ainf$-algebra $A=\bigoplus_{i,j\in\text{Ob}\mathcal A}\text{Hom}_{\mathcal A}(i,j)$ from which $\mathcal A$ can be recovered by keeping track of the sectors $\text{Hom}_{\mathcal A}(i,j)$. Therefore we can avoid the heavier notation of $\Ainf$-categories.} how they appear in the context of non-commutative geometry, and then go on to use this description to arrive at a classification result of cylic minimal models. We stress the explicit nature of this construction by phrasing it as a computer-friendly algorithm (cf.~the summary on page~\pageref{summary}). After this general discussion, in section~\ref{apptoLG} we specialise to the case of Landau-Ginzburg models and show how to apply the algorithm to matrix factorisations. Finally, we work out a few examples to recover the topological metric and compute effective superpotentials.

\section[$A_\infty$-algebras]{$\boldsymbol{A_\infty}$-algebras}\label{mathback}

In this section we discuss relevant parts of the general theory of $\Ainf$-algebras, and we explain how to construct all cyclic minimal models for a given $\Ainf$-algebra. Subsection~\ref{basicAinf} collects standard definitions and results (see also e.\,g.~\cite{g0506603,l0507222,hl0410621}) as well as detailed proofs for later use. Subsection~\ref{cycsec} offers a discussion of cyclicity and ends with an explicit step-by-step construction of cyclic minimal models.

\subsection[Basic $A_\infty$-theory]{Basic $\boldsymbol{A_\infty}$-theory}\label{basicAinf}

\begin{definition}
An \textit{$\Ainf$-algebra} $A$ is a ($\Z$- or $\Z_2$-) graded vector space together with linear maps $r_n:A[1]^{\otimes n}\rightarrow A[1]$ of degree $+1$ for all $n\geq 1$ such that
\be\label{defAinf}
\sum_{\genfrac{}{}{0pt}{}{i\geq 0,j\geq 1,}{i+j\leq n}} r_{n-j+1} \circ \left(\one^{\otimes i}\otimes r_j \otimes\one^{\otimes (n-i-j)}\right) = 0
\ee
where $A[1]$ denotes the vector space $A$ with the suspended grading, i.\,e.~if $A$ decomposes into its homogeneous components as $A=\bigoplus_iA_i$, then $A[1]_i=A_{i+1}$. 
\end{definition}
The first few $A_\infty$-conditions from~(\ref{defAinf}) read
\begin{align}\label{first3}
n=1: &\quad r_1\circ r_1 = 0 \, , \nonumber \\
n=2: &\quad r_1\circ r_2 + r_2\circ(r_1\otimes\one) + r_2\circ(\one\otimes r_1) = 0 \, , \nonumber \\
n=3: &\quad r_2\circ(r_2\otimes\one) + r_2\circ(\one\otimes r_2) \nonumber \\
& \qquad + r_1\circ r_3 + r_3\circ(r_1\otimes\one^{\otimes 2} + \one\otimes r_1\otimes\one + \one^{\otimes 2}\otimes r_1) = 0 \, ,
\end{align}
and when applied to elements in $A[1]^{\otimes n}$ these relations may pick up sign factors according to the Koszul rule, e.\,g.~$(\one\otimes r_1)(a\otimes b)=(-1)^{\widetilde a}a\otimes r_1(b)$. Here and below $\widetilde{a}$ denotes the suspended degree of $a$ in $A[1]$ which we will often simply refer to as the ``tilde degree'' to distinguish it from the degree $|a|=\widetilde a +1$ of $a$ in~$A$. 

Defining the suspension map $\sigma:A\rightarrow A[1]$ as the unique map of suspended degree $-1$ with $\sigma(a)=a$ for all $a\in A$, one may alternatively characterise $A_\infty$-algebras in terms of the maps $m_n:=\sigma^{-1}\circ r_n\circ\sigma^{\otimes n}: A^{\otimes n}\rightarrow A$. Then the relations~(\ref{first3}) say that $m_1=r_1$ is a differential with respect to the product $m_2$, this product is associative up to a homotopy given by $m_3$, and the remaining conditions in~(\ref{defAinf}) state that in general $m_n$ is associative up to a possibly non-zero homotopy $m_{n+1}$ for all $n\geq 2$. This explains the name $A_\infty$-algebra as the ``associativity up to homotopy'' may go on infinitely (though in many examples only finitely many of the higher products do not vanish). 

It follows that any differential graded (DG) algebra is in particular an $A_\infty$-algebra with $m_n=0$ for all $n\geq 3$. The reason why the products $r_n$ and not $m_n$ are used in this paper is that they reduce the amount of sign factors one has to deal with, and they also seem more natural in the reformulation in terms of non-commutative geometry to be discussed and used extensively below. 
\begin{definition}
An $A_\infty$-algebra $(A,r_n)$ is \textit{minimal} iff $r_1=0$. It is \textit{unital} iff there exists $e\in A[1]_{-1}$ such that $r_2(e\otimes a)=-a$, $r_2(a\otimes e)=(-1)^{\widetilde{a}}a$ for all $a\in A[1]$, and all other products $r_n$ vanish if applied to a tensor product involving $e$. $A$ is \textit{cyclic} with respect to a bilinear form $\langle\,\cdot\,,\,\cdot\,\rangle$ on $A$ iff
\be\label{cyclicity}
\langle a_0,r_n(a_1\otimes\ldots\otimes a_n)\rangle = (-1)^{\widetilde{a}_0(\widetilde{a}_1+\ldots+\widetilde{a}_n)} \langle a_1,r_n(a_2\otimes\ldots\otimes a_n\otimes a_0)\rangle
\ee
for all homogeneous elements $a_i\in A$.
\end{definition}
\begin{definition}
An \textit{$\Ainf$-morphism} between $\Ainf$-algebras $A$ and $A'$ is a family of linear maps $F_n:A[1]^{\otimes n}\rightarrow A'[1]$ of degree $0$ for all $n\geq 1$ such that
\be\label{defAinfmorph}
\sum_{p=1}^n \sum_{\genfrac{}{}{0pt}{}{1\leq i_1,\ldots,i_p\leq n,}{i_1+\ldots+i_p=n}} r^{A'}_p \circ \left( F_{i_1}\otimes \ldots\otimes F_{i_p}\right)= \sum_{\genfrac{}{}{0pt}{}{i\geq 0,j\geq 1,}{i+j\leq n}}\! F_{n-j+1} \circ \left(\one^{\otimes i}_A \otimes r^A_j\otimes\one^{\otimes (n-i-j)}_A\right) \, .
\ee
$(F_n)$ is an \textit{$\Ainf$-isomophism} iff $F_1$ is an isomorphism, and an \textit{$\Ainf$-quasi-isomorphism} iff $F_1$ induces an isomorphism on cohomology with respect to $r_1$. 
\end{definition}
\begin{theorem}[\cite{k0504437,m9809,ks0011041}]
Any $\Ainf$-algebra $(A,r_n)$ is $\Ainf$-quasi-isomorphic to a minimal $\Ainf$-algebra. Such a \textit{minimal model} for $A$ is unique up to $\Ainf$-isomorphisms. 
\end{theorem}
\begin{proof}
For the uniqueness property we refer to~\cite{k0504437}. To construct the minimal $\Ainf$-structure on $H_{r_1}(A)$ we adapt the proof of~\cite{m9809} to the sign conventions used in the present paper. 

Choose a vector space decomposition $A=H\oplus B\oplus L$ where $B=\text{Im}(r_1)$ and $L$ is the preimage of $B$ under $r_1$. It follows that $H\cong H_{r_1}(A)$. Now choose a homotopy map $G$ of tilde degree $-1$ such that $\one-\pi_H=r_1\circ G+G\circ r_1$ where $\pi_H$ denotes the projection to $H$. For example, one may take $G$ to be $(r_1|_L)^{-1}\circ\pi_B$. 

Next we define maps $\lambda_n: A[1]^{\otimes n}\rightarrow A[1]$ recursively by $\lambda_2:=r_2$ and
\begin{align}\label{recurminmod}
\lambda_n & := -r_2\circ(G\otimes\one)\circ(\lambda_{n-1}\otimes\one) -r_2\circ(\one\otimes G)\circ(\one\otimes\lambda_{n-1})  \nonumber \\
& \qquad - \sum_{\genfrac{}{}{0pt}{}{i,j\geq 2,}{i+j=n}} r_2\circ(G\otimes G)\circ(\lambda_i\otimes\lambda_j)
\end{align}
for all $n\geq 3$. Then one may verify that $r_n^H:=\pi_H\circ\lambda_n$ defines an $\Ainf$-structure on $H\cong H_{r_1}(A)$. This structure is related to $(A,r_n)$ by the $\Ainf$-quasi-isomorphism $(F_n):(H,r_n^H)\rightarrow(A,r_n)$ where $F_1$ is the inclusion map $H\hookrightarrow	 A$ and $F_n:=G\circ\lambda_n\circ F_1$ for $n\geq 2$. 
\end{proof}
We remark that in~\cite{l0610120} the following refinement of the above theorem is proved: if the $\Ainf$-structure on $A$ is cyclic with respect to a given pairing and if the map $G$ satisfies a certain mild cyclicity condition, then one may easily construct a minimal model on $H_{r_1}(A)$ that is cyclic with respect to the pairing induced on cohomology. This makes the computation of cyclic minimal models very straightforward in many situations. In particular this holds for the case of open topological string theory on a compact Calabi-Yau variety~$X$ in the large-volume limit, which is described by the bounded derived category of coherent sheaves on~$X$. The pairing of interest here is of course the topological metric~(\ref{DXtopmetric}) which is cyclic also off-shell. 

On the other hand, in the case of Landau-Ginzburg models the topological metric is given by~(\ref{KL}) which is cyclic only on-shell. The need to compute cyclic (and unital) minimal $\Ainf$-structures also for Landau-Ginzburg models hence calls for a more involved construction. For the rest of this section, the general theory of this construction will be explained, and in the following section it will be applied to Landau-Ginzburg models. 

\vspace{0.2cm}

It turns out that a useful equivalent description of the classical $\Ainf$-notions is in terms of the dualised bar dual, which allows for a non-commutative geometric interpretation in the sense of Kontsevich~\cite{kontsevich1993}: An $\Ainf$-algebra $(A,r_n)$ gives rise to and can be recovered from the associated formal non-commutative Q-manifold which is the tensor algebra
$$
B_A := T(A[1])^*=\bigoplus_{n\geq 0}(A[1]^*)^{\otimes n}
$$
together with a derivation $Q:B_A\rightarrow B_A$ that satisfies $Q^2=0$. If one chooses dual bases $\{e_i\}\subset A$ and $\{s^i\}\subset A[1]^*$ (which we fix from now on) then the action of the differential $Q$ on basis elements, 
$$
Q(s^a)=\sum_{n\geq 1} Q^a_{a_1\ldots a_n} s^{a_1}\otimes\ldots\otimes s^{a_n} \, ,
$$
corresponds to the classical higher products $r_n$ in such a way that both have the same coefficients $Q^a_{a_1\ldots a_n}$, 
$$
r_n(e_{a_1}\otimes\ldots\otimes e_{a_n}) = Q^a_{a_1\ldots a_n} e_a \, ,
$$
and the condition $Q^2=0$ is equivalent to the defining relations~(\ref{defAinf}) of $\Ainf$-products. Moreover, an $\Ainf$-morphism $(F_n):A\rightarrow A'$ is equivalently described by a map $F:B_{A'}\rightarrow B_A$ where the complicated relations in~(\ref{defAinfmorph}) are the same as the condition $Q\circ F=F\circ Q'$. In this language also the definition of the concatenation of two $\Ainf$-morphisms $(F_n):A_2\rightarrow A_1$ and $(G_n):A_3\rightarrow A_2$ simplifies immensely: it is simply the map $B_{A_1}\rightarrow B_{A_3}$ given by $G\circ F$.

For the time being we will adopt the point of view that not the $\Ainf$-algebra $(A,r_n)$ is the fundamental entity, but rather its dualised bar dual $(B_A,Q)$. To simplify notation we will also often write $B_A$ as $B$, $B_{A'}$ as $B'$ etc. 
\begin{definition}
The complex of \textit{non-commutative forms} over $B$ is $\Omega(B):=\bigoplus_{n\geq 0} B\otimes(B/(\C\cdot 1_B))^{\otimes n}$, where $1_B$ denotes the unit of $B$ which is the tensor product with $1\in\C$. Here the projection $B\rightarrow B/(\C\cdot 1_B)$ is denoted by $d$, and by customary abuse of notation the same symbol is also used for the differential which acts on homogeneous elements of form degree $n$ as
$$
d: b_0\otimes db_1\otimes\ldots\otimes b_n \longmapsto db_0\otimes db_1\otimes\ldots\otimes db_n \equiv 1\otimes db_0\otimes db_1\otimes\ldots\otimes db_n \, .
$$%
\end{definition}
\begin{definition}
The (\textit{de Rham} or) \textit{Karoubi complex} is given by
\be\label{Karoubi}
\mathcal{C}(B):=\Omega(B)/[\Omega(B),\Omega(B)]
\ee
together with the differential $d$ induced from $\Omega(B)$ which is again written as $d$.
\end{definition}
We denote homogeneous elements in $\mathcal{C}^n(B)$ with representatives $b_0\otimes db_1\otimes\ldots\otimes db_n$ as $(b_0\otimes db_1\otimes\ldots\otimes db_n)_{\text{c}}\equiv(b_0 db_1\ldots db_n)_{\text{c}}$ with `c' for cyclisation, and where here and from now on tensor symbols are not explicitly written if their presence is obvious from the context. The graded commutator in~(\ref{Karoubi}) is graded with respect to the combination of induced tilde degree and form degree; for example we have $(b_0db_1db_2)_{\text{c}}=-(-1)^{(\widetilde b_0+\widetilde b_1)\widetilde b_2}(db_2b_0db_1)_{\text{c}}$ where the extra minus sign comes from commuting the two differentials past each other.

\begin{proposition}[Poincar\'{e} lemma, \cite{kontsevich1993}]
$H_d^0(\mathcal{C}(B))=\C$ and $H_d^{i>0}(\mathcal{C}(B))=0$.
\end{proposition}
With these notions one can construct a complete non-commutative analogue of classical Cartan calculus. In particular, the contraction $i_\theta$ and the Lie derivative $L_\theta$ for an arbitrary derivative $\theta: B\rightarrow B$ will be relevant. They are derivatives of form degrees $-1$ and $0$, respectively, on $\Omega(B)$ (and induce suchlike derivatives on $\mathcal{C}(B)$) that are uniquely defined by $i_\theta(b)=0$, $i_\theta(db)=\theta(b)$, $L_\theta(b)=\theta(b)$ and $L_\theta(db)=d(\theta(b))$ for all $b\in B$. 

Given a morphism $\phi:B_1\rightarrow B_2$, its push-forward $\phi_*:\mathcal{C}(B_1)\rightarrow \mathcal{C}(B_2)$ is defined on homogeneous elements as $(b_0 db_1\ldots db_n)_{\text{c}} \mapsto (\phi(b_0) d(\phi(b_1))\ldots d(\phi(b_n)))_{\text{c}}$. From these definitions one may immediately verify that all the usual identities of Cartan calculus hold in the present setting, too, but we will only need the relations
\be\label{pushforcomm}
L_\theta=d\circ i_\theta + i_\theta\circ d \, ,\quad d_2\circ\phi_* = \phi_*\circ d_1 \, , \quad L_{Q_2}\circ\phi_* = \phi_*\circ L_{Q_1} \, .
\ee
\begin{definition}
A \textit{symplectic form} on $B$ is a 2-form
\be\label{symplecticform}
\omega = \omega_{ab}(ds^ads^b)_{\text{c}} + \sum_{n\geq 3}\sum_{i=1}^{n-1} \omega_{a_1\ldots a_{i};a_{i+1}\ldots a_n}\left(s^{a_1}\ldots s^{a_{i-1}}ds^{a_i}s^{a_{i+1}}\ldots s^{a_{n-1}}ds^{a_n}\right)_{\text{c}}
\ee
in $\mathcal{C}^2(B)$ such that $d\omega=0$ and $\text{det}(\omega_{ab})\neq 0$. 
\end{definition}
One can prove that the non-degeneracy condition $\text{det}(\omega_{ab})\neq 0$ is equivalent to the condition that the map $\theta\mapsto i_\theta\omega$ from derivatives on $B$ to 1-forms in $\mathcal C^1(B)$ is an isomorphism. 

The following non-commutative-geometric variant of the classical Darboux theorem will be crucial. 

\begin{theorem}[\cite{kontsevich1993,ks0606241}]
For any symplectic form $\omega\in\mathcal{C}^2(B)$ as in~(\ref{symplecticform}) there exists an automorphism $\phi:B\rightarrow B$ such that $\phi_*\omega=\omega_{ab}(ds^ads^b)_{\text{c}}$, i.\,e.~$\phi_*\omega$ is equal to the constant part of $\omega$. 
\end{theorem}
\begin{proof}
Write $\omega=\sum_{i\geq 0}\omega_i$ where $\omega_i$ has tensor degree $i+2$. The closedness condition on $\omega$ translates to its tensor components, $d\omega_i=0$ for all $i$. In particular, the existence of $\alpha_i\in\mathcal C^1(B)$ such that $\omega_1=d\alpha_1$ with $\widetilde{\omega}_1=\widetilde{\alpha}_1$ is guaranteed by the Poincar\'{e} lemma. Since $\omega_0$ is non-degenerate, there is a unique derivative $\theta_1$ that satisfies $i_{\theta_1}\omega_0=\alpha_1$ and is of tensor degree 1. It follows that $\widetilde{\theta}_1=0$ and
$$
L_{\theta_1}\omega_0=(d\circ i_{\theta_1}+i_{\theta_1}\circ d)\omega_0=d(i_{\theta_1}\omega_0)=d\alpha_1=\omega_1 \, .
$$
Now one can define a diffeomorphism $\phi_1:B\rightarrow B$ by its action on $A[1]^*$ as $\one-\theta_1$. Then the transformed symplectic form
\begin{align*}
\omega^{(1)} & := (\phi_1)_*\omega = (\phi_1)_*\omega_0 + \sum_{i\geq 1}(\phi_1)_*\omega_i \\
& = \omega_0 - \omega_{ab}\left[ (d(\theta_1(s^a))ds^b)_{\text{c}} + (ds^ad(\theta_1(s^b)))_{\text{c}}\right] \\
& \qquad + \omega_{ab}(d(\theta_1(s^a))d(\theta_1(s^b)))_{\text{c}} + \sum_{i\geq 1}(\phi_1)_*\omega_i \\
& = \omega_0-L_{\theta_1}\omega_0 + \omega_1 + \mathcal O(s^{\otimes 4}) \\
& = \omega_0 + \mathcal O(s^{\otimes 4})
\end{align*}
has no component of tensor degree 3. 

To successively transform away all higher tensor degree components, one may proceed by induction. Assume that for some $k\geq 2$ one has arrived at a symplectic form $\omega^{(k-1)}=\omega_0+\sum_{i\geq k}\omega^{(k-1)}_i$ with homogeneous tensor degree components $\omega^{(k-1)}_i$. Because of $d\omega^{(k-1)}=0$ and the Poincar\'{e} lemma, one can find $\alpha_k\in\mathcal C^1(B)$ such that $\omega^{(k-1)}_k=d\alpha_k$. This 1-form is isomorphic to a derivative $\theta_k$ of tensor degree $k$ that solves $i_{\theta_k}\omega_0=\alpha_k$, and as before it follows that $L_{\theta_k}\omega_0=\omega^{(k-1)}_k$. Then the diffeomorphism $\phi_k:B\rightarrow B$ defined on $A[1]^*$ as $\one-\theta_k$ pushes $\omega^{(k-1)}$ forward to $\omega^{(k)}:=(\phi_k)_*\omega^{(k-1)}$ which is equal to
$$
(\phi_k)_*\omega_0 + \sum_{i\geq k}(\phi_k)_*\omega^{(k-1)}_i=\omega_0-L_{\theta_k}\omega_0 + \omega^{(k-1)}_k + \mathcal O(s^{\otimes(k+3)})=\omega_0+\mathcal O(s^{\otimes(k+3)}) \, .
$$
The Darboux map $\phi$ is given by the concatenation of all $\phi_k$.
\end{proof}
To construct the Darboux map for a given symplectic form explicitly, one needs to have explicit expressions for the 1-forms $\alpha_k$ and derivatives $\theta_k$ in the above proof. Both may be read off $\omega^{(k-1)}_k$, which in general has the form
$$
\sum_{m=1}^{k+1} \omega^{(k-1)}_{a_1\ldots a_{m};a_{m+1}\ldots a_{k+2}}\left(s^{a_1}\ldots s^{a_{m-1}}ds^{a_m}s^{a_{m+1}}\ldots s^{a_{k+1}}ds^{a_{k+2}}\right)_{\text{c}} \, .
$$
The 1-form $\alpha_k$ is proportional to the contraction of $\omega^{(k-1)}_k$ with the Euler vector field $E$ which is defined as the unique derivative on $B$ that acts on elements of tensor degree $1$ as the identity, i.\,e.~$E(s^a)=s^a$. Hence one has
$$
(k+2)\omega^{(k-1)}_k=L_E \omega^{(k-1)}_k=(i_E d + di_E)\omega^{(k-1)}_k = d(i_E \omega^{(k-1)}_k)
$$
so that $\alpha_k$ may indeed be taken to be $\frac{1}{k+2}\,i_E\omega^{(k-1)}_k$.\footnote{As this reasoning works for any $n$-form with $n\geq 1$, this is essentially the proof of the Poincar\'{e} lemma.} This can be shown to be equal to
$$
\frac{2}{k+2} \sum_{m=1}^{k+1} \omega^{(k-1)}_{a_1\ldots a_{m};a_{m+1}\ldots a_{k+2}} \left(s^{a_1}\ldots s^{a_{k+1}}ds^{a_{k+2}}\right)_{\text{c}} \, ,
$$
and we see that the coefficients of $\alpha_k=\alpha_{a_1\ldots a_{k+1}b}(s^{a_1}\ldots s^{a_{k+1}}ds^{b})_{\text{c}}$ are given by
\be\label{alphacoef}
\alpha_{a_1\ldots a_{k+1}b} = \frac{2}{k+2} \sum_{m=1}^{k+1} \omega^{(k-1)}_{a_1\ldots a_{m};a_{m+1}\ldots a_{k+1} b} \, .
\ee
Next, to find an explicit expression for the derivative $\theta_k$, its action is written as $\theta_k(s^a)=\theta^a_{a_1\ldots a_{k+1}}(s^{a_1}\ldots s^{a_{k+1}})_{\text{c}}$. Then one computes
\begin{align*}
i_{\theta_k}\omega_0 & = \omega_{ab}(\theta_k(s^a)ds^b)_{\text{c}} - \omega_{ab}(ds^a\theta_k(s^b))_{\text{c}} \\
& = \omega_{ab}\theta^a_{a_1\ldots a_{k+1}}(s^{a_1}\ldots s^{a_{k+1}}ds^b)_{\text{c}} \\
& \qquad - (-1)^{\widetilde a \widetilde b + 1} (-1)^{\widetilde{a}(\widetilde{a}_1+\ldots+\widetilde{a}_{k+1})}\omega_{ba}\theta^b_{a_1\ldots a_{k+1}}(s^{a_1}\ldots s^{a_{k+1}}ds^a)_{\text{c}} \\
& = \left(1+(-1)^{2 \widetilde a \widetilde b}\right)\omega_{ab}\theta^a_{a_1\ldots a_{k+1}}(s^{a_1}\ldots s^{a_{k+1}}ds^b)_{\text{c}} \\
& = 2\omega_{ab}\theta^a_{a_1\ldots a_{k+1}}(s^{a_1}\ldots s^{a_{k+1}}ds^b)_{\text{c}} \, ,
\end{align*}
where we write the tilde degree $\widetilde{s^a}$ of $s^a$ simply as $\widetilde a$. This calculation is valid under the assumption (which is always satisfied in our applications to topological string theory) that $\omega$ is homogeneous in tilde degree, which by $\omega^{(k-1)}_k=d\alpha_k$ and $i_{\theta_k}\omega_0=\alpha_k$ implies that $\widetilde{\theta}_k=0$ and hence $\widetilde{b}=\widetilde{a}_1+\ldots+\widetilde{a}_{k+1}$ in $\theta_k(s^b)=\theta^b_{a_1\ldots a_{k+1}}(s^{a_1}\ldots s^{a_{k+1}})_{\text{c}}$ above. Therefore, the defining equation $i_{\theta_k}\omega_0=\alpha_k=\alpha_{a_1\ldots a_{k+1}b}(s^{a_1}\ldots s^{a_{k+1}}ds^b)_{\text{c}}$ for $\theta_k$ is solved if one sets
$$
\theta^c_{a_1\ldots a_{k+1}}=\frac{1}{2}\alpha_{a_1\ldots a_{k+1}b}\omega^{bc} \, .
$$
Combining this with~(\ref{alphacoef}) one arrives at the expression
\be\label{Darbouxtheta}
\theta^c_{a_1\ldots a_{k+1}} = \frac{1}{k+2} \sum_{m=1}^{k+1} \omega^{(k-1)}_{a_1\ldots a_{m};a_{m+1}\ldots a_{k+1} b} \omega^{bc}
\ee
for the components of $\theta_k$ which only depends on the constant part of $\omega$ and its recursively computed (and subsequently cancelled) higher order correction $\omega^{(k-1)}_k$.

\subsection{Cyclicity}\label{cycsec}

The following result explains how the cyclicity conditions~(\ref{cyclicity}) translate into the formulation in terms of non-commutative geometry.

\begin{proposition}
If a 2-form $\omega$ is flat, i.\,e.~$\omega=\omega_{ab}(ds^ads^b)_{\text{c}}$, then $L_Q\omega=0$ is equivalent to the cyclicity conditions
$$
\langle e_{a_0},r_n(e_{a_1}\otimes\ldots\otimes e_{a_n})\rangle = (-1)^{\widetilde{a}_0(\widetilde{a}_1+\ldots+\widetilde{a}_n)} \langle e_{a_1},r_n(e_{a_2}\otimes\ldots\otimes e_{a_n}\otimes e_{a_0})\rangle
$$
where the pairing $\langle\,\cdot\,,\,\cdot\,\rangle$ is defined via $\langle e_a,e_b\rangle=(-1)^{\widetilde{a}+1}\omega_{ab}$. 
\end{proposition}
\begin{proof}
$L_Q\omega_{ab}(ds^ads^b)_{\text{c}}$ equals
\begin{align*}
& \omega_{ab}\Big( \sum_{n\geq 1} Q_{a_1\ldots a_n}^a \sum_{i=1}^n(s^{a_1}\ldots s^{a_{i-1}}ds^{a_{i}}s^{a_{i+1}}\ldots s^{a_{n}}ds^{b})_{\text{c}} \\
& \qquad + (-1)^{\widetilde a} \sum_{n\geq 1} Q_{a_1\ldots a_n}^b (ds^a s^{a_{1}}\ldots s^{a_{i-1}} ds^{a_{i}}s^{a_{i+1}}\ldots s^{a_{n}})_{\text{c}} \Big) \\
= & \sum_{n\geq 1}\left( \omega_{ab}Q_{a_1\ldots a_n}^a - (-1)^{\widetilde b + \widetilde b(\widetilde a_1+\dots+\widetilde a_n)}\omega_{ba}Q_{a_1\ldots a_n}^a\right) (s^{a_1}\ldots s^{a_{i-1}}ds^{a_{i}}s^{a_{i+1}}\ldots s^{a_{n}}ds^{b})_{\text{c}} \\
= & \sum_{n\geq 1}2\omega_{ab} Q_{a_1\ldots a_n}^a\sum_{i=1}^n (s^{a_1}\ldots s^{a_{i-1}}ds^{a_{i}}s^{a_{i+1}}\ldots s^{a_{n}}ds^{b})_{\text{c}} \, ,
\end{align*}
where $\omega_{ab}=(-1)^{\widetilde a\widetilde b + 1}\omega_{ba}$ was used together with the fact that $Q$ is of tilde degree $+1$. Using the cyclic symmetry one now sees that $L_Q\omega$ vanishes iff
$$
\omega_{ab} Q_{a_1\ldots a_n}^a = (-1)^{(\widetilde a_1+\dots+\widetilde a_i)(\widetilde a_{i+1}+\dots+\widetilde a_n+\widetilde b)}\omega_{aa_i} Q^{a_i}_{a_{i+1}\ldots a_n b a_1\ldots a_{i-1}}
$$
for all $n\geq 1$ and all $i\in\{1,\ldots,n\}$. This holds true precisely iff all higher products $r_n$ are cyclic with respect to the pairing $\langle\,\cdot\,,\,\cdot\,\rangle$.
\end{proof}

If the 2-form $\omega$ is not flat, i.\,e.~if it has non-vanishing components of tensor order 3 or higher as in~(\ref{symplecticform}), the condition $L_Q\omega=0$ implies more complicated conditions involving further multilinear forms associated to the higher terms in $\omega$ in addition to the pairing~$\langle\,\cdot\,,\,\cdot\,\rangle$, see~\cite{l0610120}. The fact that strict cyclicity arises only for flat $\omega=\omega_{ab}(ds^ads^b)_{\text{c}}$ implies that in general an $\Ainf$-isomorphism $\phi$ will transform $\omega$ to $\phi_*\omega=\omega_{ab}(d\phi(s^a)d\phi(s^b))_{\text{c}}$ which may often not be flat. This explains why a generic minimal model will not be cyclic with respect to a given pairing: cyclicity is not an invariant under $\Ainf$-isomorphisms. 

On the other hand, the above proposition clarifies what the proper $\Ainf$-invariant generalisation of cyclicity is, namely the condition $L_Q\omega=0$ (without any further assumptions on $\omega$). Because of the last equation in~(\ref{pushforcomm}), this is indeed invariant under any $\Ainf$-morphism $\phi$ as we see from $L_{Q'}(\phi_*\omega)=\phi_*(L_Q\omega)=0$, where $Q'$ encodes the $\Ainf$-structure pushed-forward from $Q$ via $\phi$.

While cyclicity is not a natural notion in the general theory of $\Ainf$-algebras, it is still a fundamental condition in open topological string theory. However, if the 2-form $\omega$ is symplectic (as is the case in topological string theory), the Darboux theorem ensures that one can always construct an $\Ainf$-structure that is cyclic with respect to the pairing associated to the flat component of $\omega$. This is precisely what is wanted. 

It is now crucial to recognise that the development of the general theory so far allows for a more systematic study of cyclicity for a fixed $\Ainf$-algebra $A$. Instead of trying to construct $\Ainf$-maps that are cyclic with respect to a \textit{given} pairing, one may ask the more general question: What are \textit{all} the pairings on $A$ with respect to which a cyclic minimal model for $A$ exists? This question is answered by the following theorem which reformulates a result of~\cite{ks0606241}. 

\begin{theorem}
$\Ainf$-quasi-isomorphism classes in
$$
\mathscr H := H_{L_Q}(\mathcal{C}^2(B_A)_{\text{cl,hnd}})
$$
classify non-degenerate cyclic structures on minimal models of $A$ up to a change of basis. Here, $\mathcal{C}^2(B_A)_{\text{cl,hnd}}$ denotes the space of $d$-closed 2-forms that are homologically non-degenerate.  
\end{theorem}
\begin{proof}
Let $[\omega]\in\mathscr H$ with representative $\omega\in\mathcal{C}^2(B_A)_{\text{cl,hnd}}$. Further choose an arbitrary $\Ainf$-quasi-isomorphism $F$ that transports the $\Ainf$-structure on $A$ encoded in $Q$ to $H_{r_1}(A)$. By the minimal model theorem $F$ is unique up to $\Ainf$-isomorphisms. 

Because $\omega$ is $d$-closed and homologically non-degenerate, the induced form $F_*\omega$ on cohomology is symplectic. Then according to the Darboux theorem one can construct an $\Ainf$-isomorphism $\phi$ such that $\phi_*F_*\omega$ is equal to the flat part of $F_*\omega$. By assumption we have $L_Q\omega=0$, and therefore also $L_{Q'}(\phi_*F_*\omega)=0$ by~(\ref{pushforcomm}), where $Q'$ encodes the $\Ainf$-products $(r_n')$ on $H_{r_1}(A)$ pushed-forward from $Q$ via $\phi\circ F$. But the condition $L_{Q'}(\phi_*F_*\omega)=0$ is equivalent to the cyclicity of $(r_n')$ with respect to the pairing associated to the flat part of $F_*\omega$. Thus we conclude the proof by observing that 2-forms in the image of $L_Q$ can never be homologically non-degenerate. 
\end{proof}

When this result will be applied to Landau-Ginzburg models in the next section, it will turn out that in all examples we can recover the topological metric $\langle\,\cdot\,,\,\cdot\,\rangle_{\text{LG}}$ of~\cite{kl0305,hl0404} as a special case of the construction of $\mathscr H$ (any other pairing obtained this way is of course related to $\langle\,\cdot\,,\,\cdot\,\rangle_{\text{LG}}$ by a simple basis transformation). This may be viewed as an alternative derivation of the topological metric $\langle\,\cdot\,,\,\cdot\,\rangle_{\text{LG}}$ from first principles, i.\,e.~with only the defining properties of a cyclic, unital and minimal $\Ainf$-category assumed. In contrast, the derivation of $\langle\,\cdot\,,\,\cdot\,\rangle_{\text{LG}}$ in~\cite{kl0305,hl0404} relied on a path integral argument. 

To put the above theorem to practical use we need an effective method to compute the cohomology $\mathscr H$. One way to do so is to first compute $H_Q(\mathcal C^0(B)/\C)$ which is isomorphic to $H_{L_Q}(\mathcal C^2(B)_{\text{cl}})$ as we will show below. To actually obtain $\mathscr H = H_{L_Q}(\mathcal{C}^2(B_A)_{\text{cl,hnd}})$ one then has to check which elements in $H_Q(\mathcal C^0(B)/\C)$ lead to non-degenerate elements in $H_{L_Q}(\mathcal C^2(B)_{\text{cl}})$. As we will see in a moment, this second step is very simple in practice. 

In order to understand the isomorphism $H_Q(\mathcal C^0(B)/\C)\cong H_{L_Q}(\mathcal C^2(B)_{\text{cl}})$, observe that
$$
0 \longrightarrow [B,B] \stackrel{\iota}{\longrightarrow} B_+ \stackrel{\pi}{\longrightarrow} B_+/[B,B]=\mathcal C^0(B)/\C \longrightarrow 0
$$
is a short exact sequence of complexes with differentials (induced by) $L_Q$, where $\iota$ and $\pi$ are inclusion and projection maps, respectively, and $B_+:=\bigoplus_{m\geq 1}(A[1]^*)^{\otimes m}$. This gives rise to a long exact sequence in $L_Q$-cohomology, 
$$
\ldots\longrightarrow H_{L_Q}(B_+)\stackrel{\pi_*}{\longrightarrow}H_{L_Q}(B_+/[B,B])\stackrel{\delta}{\longrightarrow} H_{L_Q}([B,B])\stackrel{\iota_*}{\longrightarrow}H_{L_Q}(B_+)\longrightarrow\ldots \, ,
$$
where the connecting homomorphism $\delta$ acts as
\be\label{connecthom}
[(f)_{\text{c}}] \longmapsto [Q(f)]
\ee
and square brackets denote equivalence classes in $L_Q$-cohomology in the last expression. But according to~\cite[Prop.~7.4.1]{ks0606241}, the complex $(B_+,L_Q)$ is acylic, $H_{L_Q}(B_+)=0$, and therefore $\delta$ is an isomorphism. Because $B_+/[B,B]=\mathcal C^0(B)/\C$ by definition, and since $L_Q$ acts as $Q$ on 0-forms we really have $\delta: H_Q(\mathcal C^0(B)/\C)\cong H_{L_Q}([B,B])$.  Finally, there is another isomorphism~\cite[Prop.~5.5.1]{g0505236} between $[B,B]$ and $\mathcal C^2(B)_{\text{cl}}$ which is given by
\be\label{comm2c2}
[f,g]\longmapsto(df dg)_{\text{c}} \, .
\ee

Notice that the isomorphism $H_Q(\mathcal C^0(B)/\C)\cong H_{L_Q}(\mathcal C^2(B)_{\text{cl}})$ is completely explicit. Furthermore, it follows from the above construction that the components of elements in $H_{L_Q}(\mathcal C^2(B)_{\text{cl}})$ with lowest tensor degree are the images of the lowest tensor degree components of elements in $H_Q(\mathcal C^0(B)/\C)$. In particular, to check whether $[\omega]=[\omega_0+\omega_1+\ldots]\in H_{L_Q}(\mathcal C^2(B)_{\text{cl}})$ is non-degenerate (i.\,e.~whether $\omega_0$ is non-degenerate) one only has to consider the component $f_1$ of tensor degree $1$ of its pre-image $[(f_1+f_2+\ldots)_{\text{c}}]$. 

As a result, the main part of the computation of the space $\mathscr H$  is to compute the cohomology $H_Q(\mathcal C^0(B)/\C)$. But because of $H_Q(\mathcal{C}^0(B)/\C)\cong H_Q(\bigoplus_{m\geq 1}[(A[1]^*)^{\otimes m}]_{\text{c}})$ this can be determined using the spectral sequence~\cite{spectral} coming from the descending filtration
$$
F^n(\textstyle{\bigoplus_{m\geq 1}} [(A[1]^*)^{\otimes m}]_{\text{c}})_i = (\textstyle{\bigoplus_{m\geq n}} [(A[1]^*)^{\otimes m}]_{\text{c}})_i
$$
which converges to $H_Q(\mathcal{C}^0(B)/\C)$, and where the degree $i$ on the right-hand side is induced from the tilde grading on $A[1]$.

A direct calculation shows that this spectral sequence simply computes $H_Q(\mathcal C^0(B)/\C)$ tensor order by tensor order, i.\,e.~the sum over the $r$-th terms in the spectral sequence is equal to $H_Q(\mathcal C^0(B)/\C)$ up to elements of tensor degree $r+1$ or higher. In the next section we will see an explicit example of such calculations. 

\subsubsection*{Summary: explicit construction of cyclic minimal models}\label{summary}

In conclusion, starting from an arbitrary $\Ainf$-algebra $(A,r_n)$ we have seen in this section how to find all non-degenerate pairings on $H=H_{r_1}(A)$ with respect to which cyclic minimal models of $(A,r_n)$ exist, and furthermore these cyclic $\Ainf$-structures on $H$ can be explicitly constructed. 

The case of interest for open topological string theory is when $(A,r_n)$ is an off-shell DG algebra, i.\,e.~$r_n=0$ for all $n\geq 3$.\footnote{This is not really a special case as there exists an ``anti-minimal model'' theorem~\cite{l0310337}: Any $\Ainf$-algebra is $\Ainf$-quasi-isomorphic to a DG algebra.} For this case we summarise the explicit construction of cyclic minimal models for $A$ as follows. 
\begin{enumerate}
\item Compute $H_Q(\mathcal C^0(B_A)/\C)$: to do this, write $Q=Q_1+Q_2$ with $Q_i$ dual to $r_i$ and recursively solve the equations
$$
Q_1(f_1) = 0 \, , \quad Q_2(f_i)_{\text{c}} = - Q_1(f_{i+1})_{\text{c}} \, , \quad i\geq 1 \, ,
$$
for $f_i\in(A[1]^*)^{\otimes i}$. Then $[(f_1+f_2+\ldots)_{\text{c}}]$ is non-trivial in $H_Q(\mathcal C^0(B_A)/\C)$. 
\item Compute $\mathcal C^2(B_A)_{\text{cl}}$ using the isomorphisms~(\ref{connecthom}),~(\ref{comm2c2}):
$$
H_Q(\mathcal C^0(B_A)/\C) \ni [(f_1+f_2+\ldots)_{\text{c}}] \longmapsto [(\omega_0+\omega_1+\ldots)_{\text{c}}] \in H_{L_Q}(\mathcal C^2(B_A)_{\text{cl}}) \, .
$$
\item Obtain $\mathscr H = H_{L_Q}(\mathcal{C}^2(B_A)_{\text{cl,hnd}})$ by discarding those elements $[\omega]=[(\omega_0+\omega_1+\ldots)_{\text{c}}]\in H_{L_Q}(\mathcal C^2(B_A)_{\text{cl}})$ for which the matrix $(\omega_{ab})$ in $\omega_0=\omega_{ab}(ds^ads^b)_{\text{c}}$ is not invertible when restricted to $r_1$-cohomology.
\item Construct an arbitrary (possibly non-cyclic) minimal model $(H=H_{r_1}(A),r_n')$ with $\Ainf$-quasi-isomorphism $(F'_n): H \rightarrow A$ using~(\ref{recurminmod}). 
\item Compute the symplectic form $F'_*\omega\in\mathcal C^2(B_H)$. 
\item Construct the Darboux map $\phi=\prod_i(\one-\theta_i)$ from~(\ref{Darbouxtheta}) as the symplectomorphism $(B_H,Q',F'_*\omega)\rightarrow (B_H,Q_{\text{min}},\phi_*F'_*\omega)$ where $Q_{\text{min}}$ encodes the $\Ainf$-structure pushed-forward from $Q'$ via $\phi$.
\item Obtain the $\Ainf$-products $r_n^{\text{min}}$ pertaining to $Q_{\text{min}}$ from~(\ref{defAinfmorph}), i.\,e.
\begin{align*}
r_n^{\text{min}} & = \sum_{\genfrac{}{}{0pt}{}{i\geq 0,j\geq 1,}{i+j\leq n}}\! \phi_{n-j+1} \circ \left(\one^{\otimes i}_H \otimes r'_j\otimes\one^{\otimes (n-i-j)}_H\right) \\
& \qquad - \sum_{p=1}^{n-1} \sum_{\genfrac{}{}{0pt}{}{1\leq i_1,\ldots,i_p\leq n,}{i_1+\ldots+i_p=n}} r_p^{\text{min}} \circ \left( \phi_{i_1}\otimes \ldots\otimes \phi_{i_p}\right) \, .
\end{align*}
By construction, the higher products $r_n^{\text{min}}$ are cyclic with respect to the pairing defined by $\langle e_a,e_b\rangle=(-1)^{\widetilde a + 1}\omega_{ab}$.
\end{enumerate}

The above algorithm can be implemented universally on a computer to construct cyclic minimal models for any DG algebra (independent of whether it is $\Z_2$- or $\Z$-graded). The only input necessary is the algebraic structure in terms of the numbers $Q^a_b$, $Q^a_{bc}$. Then if $H_Q(\mathcal C^0(B_A)/\C)$ is computed up to tensor degree $N$ in step (i), the algorithm produces a minimal model which is guaranteed to be cyclic up to order $N+1$. While elements in $H_Q(\mathcal C^0(B_A)/\C)$ will typically be infinite sums, the cyclic $\Ainf$-products $r_n^{\text{min}}$ will often vanish for sufficiently large $n$. In this case the algorithm produces a full cyclic minimal model after a finite number of steps.

\vspace{0.2cm}

We close this section with a remark that will not be relevant for the rest of the paper. As discussed above, the approach here is to find all cyclic pairings and then recover the one of interest in the concrete application. For the case of Landau-Ginzburg models one may also try to directly find minimal $\Ainf$-products that are cyclic with respect to the topological metric $\langle\,\cdot\,,\,\cdot\,\rangle_{\text{LG}}$. By the Darboux theorem and the non-commutative-geometric characterisation of cyclicity this amounts to finding higher order terms such that
$$
\omega = \omega_{ab}(ds^ads^b)_{\text{c}} + \sum_{n\geq 3}\sum_{i=1}^{n-1} \omega_{a_1\ldots a_{i};a_{i+1}\ldots a_n}\left(s^{a_1}\ldots s^{a_{i-1}}ds^{a_i}s^{a_{i+1}}\ldots s^{a_{n-1}}ds^{a_n}\right)_{\text{c}}
$$
with $\omega_{ab}=(-1)^{\widetilde a + 1}\langle e_a,e_b\rangle_{\text{LG}}$ is both $d$-closed an $L_Q$-closed. As explained in~\cite{l0610120} the latter condition is equivalent to the existence of multilinear maps $\langle\,\cdot\,,\ldots,\,\cdot\,\rangle_{i,n}:A^{\otimes n}\rightarrow\C$ corresponding to $\omega_{a_1\ldots a_{i};a_{i+1}\ldots a_n}$ that obey certain compatibility conditions with the DG structure on $A$. Let $D$ be a matrix factorisation describing a brane in a Landau-Ginzburg model with potential $W$ in $N$ chiral fields. Then by direct computation one can verify that in this case the multilinear forms defined by
\begin{align*}
& \left\langle e_{a_1}, \ldots, e_{a_n} \right\rangle_{1,n}^D \\
= & \oint \frac{-\D x_1\ldots \D x_N}{(2\pi\I)^N\prod_{i=1}^N \partial_i W} \text{str}\bigg(\sum_{i_1=1}^{N-n+3} \sum_{i_2=i_1 +1}^{N-n+4} \ldots \sum_{i_{n-2}=i_{n-3}+1}^N \\
& \quad \cdot (-1)^{(N+n)|e_{a_1}| + \sum_{j=1}^{n-2}(i_j+\varepsilon_j^{[n]})|e_{a_{j+1}}| + \sum_{j=1}^{n-2}i_j + \varepsilon^{[n]}} \\
& \quad \cdot e_{a_1} \frac{\partial D}{\partial x_1}\ldots \frac{\partial D}{\partial x_{i_1 -1}} \frac{\partial e_{a_2}}{\partial x_{i_1}} \frac{\partial D}{\partial x_{i_1 +1}} \ldots\;\ldots \frac{\partial D}{\partial x_{i_{n-2} -1}} \frac{\partial e_{a_{n-1}}}{\partial x_{i_{n-2}}} \frac{\partial D}{\partial x_{i_{n-2} +1}} \ldots \frac{\partial D}{\partial x_{N}} e_{a_n} \bigg)
\end{align*}
with $\varepsilon^{[n]}_j:=N+n+j$, $\varepsilon^{[1]}:=N$ and $\varepsilon^{[n+1]}:=\varepsilon^{[n]}+N+n+1$, give rise to higher corrections to $\omega_{ab}(ds^ads^b)_{\text{c}}$ such that $L_Q\omega=0$ holds off-shell. Moreover, in explicit examples one can check that the thus constructed $\omega$ is also $d$-closed for many choices of branes~$D$, but not for every brane; it is unclear what general property of~$D$ may prevent the above~$\omega$ from being $d$-closed in the latter case. If $d\omega\neq 0$, additional corrections to $\omega_{ab}(ds^ads^b)_{\text{c}}$ would have to be found, while in the former case one can immediately construct cyclic minimal models without having to compute $H_Q(\mathcal C^0(B_A)/\C)$ first. However, in the present paper we only use the construction detailed in the above algorithm as it is much more generally applicable.

\section{Application to Landau-Ginzburg models}\label{apptoLG}

We will now apply the results of the previous section to construct cyclic, unital and minimal $\Ainf$-products for Landau-Ginzburg models. This establishes explicitly the full structure of open topological string theory for such models, and it allows to algorithmically compute effective superpotentials. 

D-brane systems in twisted Landau-Ginzburg models with superpotential $W$ are described by matrix factorisations $D$ of $W$, and on-shell open string states of such branes correspond to the cohomology of the BRST operator $[D,\,\cdot\,]$. We are interested in endowing BRST cohomology with the proper cyclic, unital and minimal $\Ainf$-structure. Hence, for a given matrix factorisations $D$ (which may of course correspond to an arbitrary superposition of branes) of rank $r$, we set\footnote{Recall that the relation between the products $r_n$ and $m_n=\sigma^{-1}\circ r_n\circ\sigma^{\otimes n}$ was explained at the beginning of section~\ref{basicAinf}.}
$$
A:=\text{Mat}(\C[X],2r) \, , \quad m_1:=[D,\,\cdot\,] \, , \quad m_2:=\text{matrix multiplication.}
$$
This is the off-shell DG algebra to which the construction explained in the previous section can be applied to find all cyclic $\Ainf$-structures on cohomology, i.\,e.~on the boundary chiral ring. 

\vspace{0.2cm}

We will start off lightly in subsection~\ref{app1} where we will carry out only steps~(i)--(iii), and only to first order, of the algorithm in subsection~\ref{cycsec} for a number of examples. This will allow us to see how the topological metric of~\cite{kl0305,hl0404} may be recovered from the systematic approach followed in the present paper. Then we will give details on how to carry out step~(i) to all orders for A-type minimal conformal models, i.\,e.~how to compute $H_Q(\mathcal C^0(B)/\C)$. Finally, in subsection~\ref{app2} examples of the calculation of all amplitudes and effective superpotentials will be presented by executing the full algorithm. 

\subsection{First examples}\label{app1}

\noindent\textbf{Example 1: transposition branes. } We consider the Landau-Ginzburg potential $x^n+y^n$ and its matrix factorisation $D=(\begin{smallmatrix}0&(x^n+y^n)/(x-\eta y)\\ x-\eta y&0\end{smallmatrix})$, where $\eta$ is an $n$-th root of $-1$, see~\cite{add0401}. Then BRST cohomology is simply given by its basis representatives $e_i:=(\begin{smallmatrix}y^i&0\\ 0&y^i\end{smallmatrix})$ for $i\in\{0,\ldots,n-2\}$. Denoting the dual basis by $s^i$ as usual, step~(i) is trivial to first tensor order since $Q_1(s^i)=0$ is the same as $r_1(e_i)=[D,e_i]=0$. 

To carry out step~(ii) one has to know the action of $Q_2$, or dually the full multiplication structure of the off-shell DG algebra $A$. However, we are at the moment only interested to compute in first tensor order, and from $e_i=e_k e_{i-k}=e_{i-k}e_k$ we immediately see that
$$
Q(s^i)=Q_2(s^i)=-\sum_{k=0}^i s^k s^{i-k} =
\begin{cases}
-{\displaystyle\sum_{k=0}^{i/2-1}}[s^k,s^{i-k}] - \frac{1}{2}[s^{i/2},s^{i/2}] + \ldots \\
-{\displaystyle\sum_{k=0}^{(i+1)/2}}[s^k,s^{i-k}] + \ldots
\end{cases}
$$
where the two cases are for $i$ even or odd, respectively, and `$+\ldots$' denotes the contribution from basis elements other than $s^k$, i.\,e.~from elements that are dual not to $e_k$ but to the remaining basis elements of $A$. These contributions can straightforwardly be calculated, but they are not relevant in first order. Thus we have already computed the map~(\ref{connecthom}) from $H_Q(\mathcal C^0(B)/\C)$ to $H_Q([B,B])$ in step~(ii). To complete this step we apply the isomorphism~(\ref{comm2c2}) and find that the flat part of the form $\omega=\omega_{ab}(ds^ads^b)_{\text{c}}\in\mathcal C^2(B_{H_{r_1}(A)})$ is given by
\be\label{omegalambda}
(\omega_{ab}) =
\begin{pmatrix}
\lambda_0&\lambda_1&\cdots&\lambda_{n-3}&\lambda_{n-2}\\
\lambda_1&&\iddots&\iddots\\
\vdots&\iddots&\iddots\\
\lambda_{n-3}&\iddots\\
\lambda_{n-2}\\
\end{pmatrix}
\ee
with arbitrary complex numbers $\lambda_i$. Step~(iii) of the algorithm is now simply to note that this matrix is non-degenerate (and therefore gives rise to a symplectic form~$\omega$) iff~$\lambda_{n-2}\neq 0$. Furthermore by setting $\lambda_0=\ldots=\lambda_{n-3}=0$ we precisely recover the on-shell topological metric $\langle\,\cdot\,,\,\cdot\,\rangle_{\text{LG}}$ in this example. 

\vspace{0.2cm}

\noindent\textbf{Example 2: linear matrix factorisations. } Let us next work out the case of so-called linear matrix factorisations of the cubic Landau-Ginzburg potential $W=x_1^3+x_2^3+x_3^3$. As explained in~\cite{err0508}, for any third root $\eta$ of $-1$ there are matrices
\begin{align*}
\alpha_0 & = \begin{pmatrix}
x_{1} & x_{2}-\eta x_{3} & 0 \\
0 & x_{1} & x_{2}+x_{3} \\
x_{2}+(\eta -1)x_{3} & 0 & x_{1}
\end{pmatrix} \\
\alpha_1 & = \begin{pmatrix}
x_{1} & (\eta -1)x_{2}+x_{3} & 0 \\
0 & x_{1} & (\eta -1)x_{2}+(\eta -1)x_{3} \\
(\eta -1)x_{2}-\eta x_{3} & 0 & x_{1}
\end{pmatrix} \\
\alpha_2 & = \begin{pmatrix}
x_{1} & -\eta x_{2}+(\eta -1)x_{3} & 0 \\
0 & x_{1} & -\eta x_{2}-\eta x_{3} \\
-\eta x_{2}+x_{3} & 0 & x_{1}
\end{pmatrix}
\end{align*}
with the property that $\alpha_{\sigma(0)}\alpha_{\sigma(1)}\alpha_{\sigma(2)}=W\one$ for any permutation $\sigma\in S_3$. This gives rise to a matrix factorisation $D=(\begin{smallmatrix}0&\alpha_2\\ \alpha_0 \alpha_1&0\end{smallmatrix})$. Now we apply the method of~\cite{TCinTopDBcat} to compute explicit basis representatives
of BRST cohomology $H$. The result is that the even and odd subspaces of $H$ are both four-dimensional and we denote their basis elements $e_1,\ldots,e_8$ (we do not display the explicit matrices here, but they are included in the tex-file of this document). Upon working out the multiplication structure of $H$ we find that
$$
Q_2(s^5) = -[s^1,s^5] + \frac{\eta+1}{\eta}[s^2,s^6] - \frac{2\eta-1}{\eta}[s^3,s^7] - \eta [s^4,s^8] - [s^4,s^6] + \ldots \, .
$$
It follows that the flat part of the 2-form $\omega\in\mathcal C^2(B_H)$ corresponding to $(s^5+\ldots)_{\text{c}}\in\mathcal C^0(B_A)/\C$ is given by
$$
(\omega_{ab}) =
\begin{pmatrix}
&&&&-1&0&0&0 \\
&&&&0&\frac{\eta+1}{\eta}&0&0 \\
&&&&0&0&-\frac{2\eta-1}{\eta}&0 \\
&&&&0&-1&0&-\eta \\
1&0&0&0 \\
0&-\frac{\eta+1}{\eta}&0&1 \\
0&0&\frac{2\eta-1}{\eta}&0 \\
0&0&0&\eta
\end{pmatrix}.
$$
It is easy to verify that $(-1)^{\widetilde a+1}\omega_{ab}=(1+\frac{5}{1+3\eta^2})\langle e_a,e_b\rangle_{\text{LG}}$ and hence we again recover the topological metric. 

\vspace{0.2cm}

\noindent\textbf{Example 3: minimal conformal models. } As a third example we consider arbitrary branes in Landau-Ginzburg models with potential $-x^n$. As any matrix factorisation of $-x^n$ is isomorphic to a direct sum of the simple factorisations $D_a:=(\begin{smallmatrix}0&x^{n-a}\\-x^a&0\end{smallmatrix})$, it is sufficient to only work out the case of the matrix factorisations $D_a\oplus D_b$ for any $a,b\in\{1,\ldots,n-1\}$. 

If we choose BRST cohomology to be represented as
\begin{align*}
H & = H_a^0 \oplus H_a^1 \oplus H_{ab}^0 \oplus H_{ab}^1 \oplus H_{ba}^0 \oplus H_{ba}^1 \oplus H_{b}^0 \oplus H_{b}^1\,, \\
H_a^0 & = \C\left\{ e^{(a,+)}_i = \begin{pmatrix}x^i&0\\ 0&x^i\end{pmatrix} \Big|\; 0\leq i\leq\min(a-1,n-a-1) \right\}\,, \\
H_a^1 & = \C\left\{ e^{(a,-)}_i = \begin{pmatrix}0&x^{n-2a+i}\\ x^i&0\end{pmatrix} \Big|\; \max(2a-n,0)\leq i\leq a-1 \right\}\,, \\
H_{ab}^0 & = \C\left\{ e^{(ab,+)}_i = x^i\begin{pmatrix}x^{a-b}&0\\ 0&1\end{pmatrix} \Big|\; \max(b-a,0)\leq i\leq \min(b-1,n-a-1) \right\}\,, \\
H_{ab}^1 & = \C\left\{ e^{(ab,-)}_i = x^i\begin{pmatrix}0&x^{n-a-b}\\ 1&0\end{pmatrix} \Big|\; \max(a+b-n,0)\leq i\leq \min(a-1,b-1) \right\}
\end{align*}
and define $d_a=\text{dim}H_a^0$, $d_{ab}=\text{dim}H_{ab}^0$, it is straightforward to compute the action of $Q_2$ on the dual basis elements. In particular, one finds that $Q_2(s^{a-1}_{(a,-)} + s^{b-1}_{(b,-)})$ is equal to
\begin{align}\label{Q2onsum}
& \sum_{i=1}^{d_a}\left[s_{(a,-)}^{a-i}, s_{(a,+)}^i\right] + \sum_{j=1}^{d_b}\left[s_{(b,-)}^{b-j}, s_{(b,+)}^j\right] \nonumber \\
& \quad + \sum_{k=0}^{d_{ab}-1}\left( \left[s_{(ab,-)}^{\min(a-1,b-1)-k}, s_{(ba,+)}^{\max(a-b,0)+k}\right] + \left[s_{(ba,-)}^{\max(a+b-n,0)+k}, s_{(ab,+)}^{\min(b-1,n-a-1)-k}\right]\right) \nonumber \\
& \quad + \ldots
\end{align}
where `$+\ldots$' again denotes contributions that do not affect the first order result. From~(\ref{Q2onsum}) one sees that the flat part of the associated symplectic form on cohomology is given by
$$
\left(\begin{array}{c@{}c@{}c@{}c}
\begin{array}{|ccc|}\hline
0&&1\\
&\iddots&\\
-1&&0\\\hline
\end{array}&0&0&0\\
0&0&\begin{array}{|ccc|}\hline
0&&1\\
&\iddots&\\
-1&&0\\\hline
\end{array}&0\\
0&\begin{array}{|ccc|}\hline
0&&1\\
&\iddots&\\
-1&&0\\\hline
\end{array}&0&0\\
0&0&0&\begin{array}{|ccc|}\hline
0&&1\\
&\iddots&\\
-1&&0\\\hline
\end{array}
\end{array}\right)
$$
where the upper-left and lower-right blocks have sizes $2d_a$ and $2d_b$, respectively, and the two blocks in the middle both have size $2d_{ab}$. Again, this is the form the topological metric $\langle\,\cdot\,,\,\cdot\,\rangle_{\text{LG}}$ takes in this choice of basis. If we do not start out with the element $[s^{a-1}_{(a,-)} + s^{b-1}_{(b,-)}+\ldots]\in H_Q(\mathcal C^0(B)/\C)$ but more generally with $[\sum_j\lambda_j(s^{a-1-j}_{(a,-)} + s^{b-1-j}_{(b,-)})+\ldots]$, we obtain a symplectic form whose blocks are similar to~(\ref{omegalambda}). 

\vspace{0.2cm}

We remark that all of the examples studied -- most of which are not described here for brevity -- allow to recover the topological metric $\langle\,\cdot\,,\,\cdot\,\rangle_{\text{LG}}$ in the general first principle approach followed here. Also, all examples share the feature that the flat part of the symplectic form always comes from one single element $[s^a+\ldots]\in H_Q(\mathcal C^0(B)/\C)$, where $s^a$ is dual to the basis element $e_a$ of highest polynomial degree. If there are fermionic elements in the spectrum, $e_a$ is always one of those.

\subsubsection*{Computing $\boldsymbol{H_Q(\mathcal C^0(B)/\C)}$}

So far we have only computed the space $H_Q(\mathcal C^0(B)/\C)$ in step~(i) to first order, but to compute amplitudes one needs more than that. In concrete examples we will use a computer implementation of the ``perturbative'' calculation of $H_Q(\mathcal C^0(B)/\C)$ and all the other steps of our algorithm. Before this will be discussed in the next subsection, we will now illustrate the computation of $H_Q(\mathcal C^0(B)/\C)$ in the case of A-type minimal conformal models. 

We consider the matrix factorisation $D_a=(\begin{smallmatrix}0&x^{n-a}\\-x^a&0\end{smallmatrix})$ of $-x^n$. Then if $a\geq n$ a basis for the off-shell DG algebra $A=\text{Mat}(\C[x],2)$ is given by
\begin{align}\label{Abasis}
e_i^{(+)} & =
\begin{pmatrix}
x^i & 0\\ 0 & x^i
\end{pmatrix}
, \quad
&& e_i^{(L,+)}  =
\begin{pmatrix}
0 & 0\\ 0 & x^i
\end{pmatrix}
, \quad \nonumber \\
e_i^{(-)} & =
\begin{pmatrix}
0 & x^i\\ -x^{2a-n+i} & 0
\end{pmatrix}
, \quad
&& e_i^{(L,-)} =
\begin{pmatrix}
0 & 0\\ x^i & 0
\end{pmatrix}
\end{align}
with $i\geq 0$.\footnote{In the case $a\leq n-a$ one may instead choose the basis $(\begin{smallmatrix}x^i & 0\\ 0 & x^i\end{smallmatrix}),  (\begin{smallmatrix}0 & -x^{2a-n+i} \\ x^i & 0\end{smallmatrix}), (\begin{smallmatrix}x^i & 0\\ 0 & 0\end{smallmatrix}), (\begin{smallmatrix}0 & x^i\\ 0 & 0\end{smallmatrix})$, and the subsequent argument will apply in the exact same way.} We also define $e_i^{(B,\pm)}= e_{n-a-1+i}^{(\pm)}$ as a basis for the image of $r_1=[D_a,\,\cdot\,]$ and note that $\C\{e_i^{(\pm)}\}_{i\in\{0,\ldots,n-a-1\}}\cong H_{r_1}(A)$. 

From example 3 above we know that the element $s_{(-)}^{n-a-1}$ dual to $e_{n-a-1}^{(-)}$ corresponds to the topological metric. Therefore, we now wish to construct an element $[f]$ in $H_Q(\mathcal C^0(B)/\C)$ whose component of tensor order 1 is equal to $s_{(-)}^{n-a-1}$. For this it will be sufficient to know that $Q_1(s_{(B,\pm)}^i)=s_{(L,\mp)}^{i-1}$ and
\be\label{Q2sminusi}
Q_2(s_{(-)}^i) = \sum_{j=0}^i\left(s_{(-)}^i s_{(+)}^{i-j} - s_{(+)}^{i-j} s_{(-)}^{j} + s_{(-)}^{j} s_{(L,+)}^{i-j}\right) \, .
\ee
These actions follow directly from the explicit choice of basis~(\ref{Abasis}). From~(\ref{Q2sminusi}) we see that $f_1:=s_{(-)}^{n-a-1}$ is a $Q$-cohomology representative only to first tensor order: 
$$
Q(s_{(-)}^{n-a-1})_{\text{c}}=Q_2(s_{(-)}^{n-a-1})_{\text{c}}=\sum_{j=0}^{n-a-1}(s_{(-)}^{j}s_{(L,+)}^{n-a-1-j})_{\text{c}} \, .
$$
To obtain a representative of $Q$-cohomology also to second tensor order, one observes that
$$
\sum_{j=0}^{n-a-1} s_{(-)}^{j}s_{(L,+)}^{n-a-1-j} = Q_1\Big(\sum_{j=0}^{n-a-1}s_{(-)}^{j}s_{(B,-)}^{n-a-j}\Big) =: -Q_1(f_2)
$$
and hence $Q=Q_1+Q_2$ acting on $f_1+f_2$ gives zero up to the term $-Q_2(\sum_{j=0}^{n-a-1}s_{(-)}^{j}s_{(B,-)}^{n-a-j})$, which is of tensor degree 3. A short calculation shows that this term is equal to the $Q_1$-image of
$$
-f_3 := -\sum_{j_1=0}^{n-a-1}\sum_{j_2=0}^{j_1}(s_{(-)}^{j_2} s_{(B,-)}^{j_1-j_2+1}s_{(B,-)}^{n-a-j_1})_{\text{c}} -\sum_{j_1=0}^{n-a-1}\sum_{j_2=0}^{n-a-1}(s_{(-)}^{j_1} s_{(-)}^{j_2}s_{(B,-)}^{2n-2a-j_1-j_2})_{\text{c}} \, .
$$
Thus $f_1+f_2+f_3$ is a $Q$-cohomology representative up to tensor order 3. Increasingly tedious computations reveal that this correction process, one tensor order at a time, can be continued. For example, the next order is given by
\begin{align*}
f_4 = & \sum_{j_1=0}^{n-a-1}\sum_{j_2=0}^{j_1}\sum_{j_3=0}^{j_2}(s_{(-)}^{j_3} s_{(B,-)}^{j_2-j_3+1}s_{(B,-)}^{j_1-j_2+1}s_{(B,-)}^{n-a-j_1})_{\text{c}} \\
& \quad + \sum_{j_1=0}^{n-a-1}\sum_{j_2=0}^{n-a-1}\sum_{j_3=0}^{2n-2a-j_1-j_2-1}(s_{(-)}^{j_1} s_{(-)}^{j_2} s_{(B,-)}^{j_3+1} s_{(B,-)}^{2n-2a-j_1-j_2-j_3})_{\text{c}} \\
& \quad + \sum_{j_1=0}^{n-a-1}\sum_{j_2=0}^{n-a-1}\sum_{j_3=0}^{j_1}(s_{(-)}^{j_3} s_{(B,-)}^{j_1-j_3+1} s_{(-)}^{j_2} s_{(B,-)}^{2n-2a-j_1-j_2})_{\text{c}} \\
& \quad + \sum_{j_1=0}^{n-a-1}\sum_{j_2=0}^{n-a-1}\sum_{j_3=0}^{n-a-1}(s_{(-)}^{j_1} s_{(-)}^{j_2} s_{(-)}^{j_3} s_{(B,-)}^{3n-3a-j_1-j_2-j_3})_{\text{c}} \, .
\end{align*}
One may now go on to identify the general structure of these tensor components and obtain all higher order corrections to $s_{(-)}^{n-a-1}$, and also for $s_{(-)}^{n-a-1-j}$ for all $j\in\{1,\ldots,n-a-1\}$. Similar computations, but somewhat more involved and heavier in notation, allow to construct $H_Q(\mathcal C^0(B)/\C)$ also for multiple brane system such as the superposition $D_a\oplus D_b$ of example~3. This then concludes the hardest part of our algorithm to construct cyclic, unital and minimal $\Ainf$-products for any object in $\text{MF}(x^n)$. However, we refrain from providing more technical details here and instead now go on to use the full algorithm to compute effective superpotentials.

\subsection{Amplitudes and effective superpotentials}\label{app2}

The algorithm of subsection~\ref{cycsec} can be implemented on a computer to construct all cyclic minimal models for any DG algebra $A$ if its structure constants are provided as input. In the case of Landau-Ginzburg models not even that is necessary, it suffices to specify the matrix factorisation $D$ describing the boundary sector of interest. We then add a step~(0) to our algorithm which computes an explicit basis of $[D,\,\cdot\,]$-cohomology as explained in~\cite{TCinTopDBcat} and furthermore constructs from this an off-shell basis up to a given polynomial degree. This basis can always be chosen to be compatible with the decomposition $A=H\oplus B\oplus L$ as in the proof of the minimal model theorem. In particular, this also allows to construct the homotopy or propagator $G$ as an ``inverse'' to $[D,\,\cdot\,]$ for any matrix factorisation in any Landau-Ginzburg model. We also note that the general construction of cyclic minimal models applies equally well to $\Z_2$- and $\Z$-graded matrix factorisation. To compute the examples in this section the complete algorithm has been implemented using the computer algebra system Singular~\cite{Singular}. 

\vspace{0.2cm}

The first example that we discuss in some detail to illustrate the procedure is that of the matrix factorisation $D=(\begin{smallmatrix}0&x^2\\x^3&0\end{smallmatrix})$. The effective superpotential $\mathcal W_{\text{eff}}$ pertaining to this brane has been computed by a different method in~\cite{hll0402}, so we will have something to compare our result to. 

For the basis of BRST cohomology $H$ we make the choice
$$
e_1=
\begin{pmatrix}
1&0\\0&1
\end{pmatrix}, \quad
e_2=
\begin{pmatrix}
x&0\\0&x
\end{pmatrix}, \quad
e_3=
\begin{pmatrix}
0&-1\\x&0
\end{pmatrix}, \quad
e_4=
\begin{pmatrix}
0&-x\\x^2&0
\end{pmatrix}.
$$
As we know from our previous discussion, we now want to compute a flat symplectic form on BRST cohomology that comes from an element of $\mathcal C^0(B_A)/\C$ whose first tensor component is dual to the fermionic open string state $e_4$. Feeding this data into our algorithm and computing up to tensor order 4, the result of steps (i)--(iii) is a 2-form $\omega\in\mathcal C^2(B_A)$ which is a sum of $1089$ terms. Those terms of the flat part of $\omega$ that will survive the push-forward to $H$ in step~(v) are simply given by
\be\label{dd}
\omega_{ab}(ds^ads^b)_{\text{c}} \, , \quad
(\omega_{ab}) =
\begin{pmatrix}
0&0&0&-1\\
0&0&-1&0\\
0&1&0&0\\
1&0&0&0
\end{pmatrix}
\ee
and this of course corresponds precisely to the topological metric $\langle\,\cdot\,,\,\cdot\,\rangle_{\text{LG}}$. We will not write out the remaining $1087$ terms of $\omega$ here. Instead, we go on to step~(iv) and compute a minimal model with $\Ainf$-products $r_n'$ on $H$. The non-vanishing coefficients are
\begin{align}\label{noncycAinf}
& {Q'}_{11}^1 = -1\, , \quad && {Q'}_{223}^1 = -1\, , \quad && {Q'}_{2244}^1 = +1\, , \quad && {Q'}_{44444}^1 = -1\, . \nonumber\\
& {Q'}_{12}^2 = -1\, , \quad && {Q'}_{224}^2 = -1\, , \quad && {Q'}_{2444}^3 = +1\, , \quad && \nonumber\\
& {Q'}_{13}^3 = -1\, , \quad && {Q'}_{234}^3 = -1\, , \quad && {Q'}_{3444}^1 = +1\, , \quad && \nonumber\\
& {Q'}_{14}^4 = -1\, , \quad && {Q'}_{243}^3 = -1\, , \quad && {Q'}_{4344}^1 = +1\, , \quad && \nonumber\\
& {Q'}_{21}^2 = -1\, , \quad && {Q'}_{324}^3 = +1\, , \quad && {Q'}_{4442}^3 = -1\, , \quad && \nonumber\\
& {Q'}_{23}^4 = -1\, , \quad && {Q'}_{343}^1 = -1\, , \quad && {Q'}_{4443}^1 = +1\, , \quad && \nonumber\\
& {Q'}_{31}^3 = +1\, , \quad && {Q'}_{424}^4 = +1\, , \quad && {Q'}_{4444}^2 = +1\, , \quad && \nonumber\\
& {Q'}_{32}^4 = +1\, , \quad && {Q'}_{432}^3 = +1\, , \quad &&  \quad && \nonumber\\
& {Q'}_{33}^2 = -1\, , \quad && {Q'}_{433}^1 = -1\, , \quad &&  \quad && \nonumber\\
& {Q'}_{41}^4 = +1\, , \quad && {Q'}_{434}^2 = -1\, , \quad &&  \quad && \nonumber\\
&   && {Q'}_{442}^4 = -1\, , \quad &&  \quad &&
\end{align}
Comparing this with~(\ref{dd}) we realise that the $\Ainf$-structure $(r_n')$ is not cyclic. As explained in the previous section, this is generically expected and indeed the main point of the present paper. 

The non-cyclicity of the products in~(\ref{noncycAinf}) is consistent with the fact that the push-forward $F'_*\omega$ of $\omega$ under the $\Ainf$-quasi-isomorphism $(F'_n): H\rightarrow A$ obtained in step~(v) is not flat.\footnote{We do not give $F'$ explicitly here because this would also make it necessary to explicify our choice of the (rather large) off-shell basis.} On the other hand, the symplectic form $\phi_*F'_*\omega$ which we obtain from the construction of the Darboux map $\phi$ in step~(vi) \textit{is} flat (and of course equal to~(\ref{dd})). The non-vanishing coefficients of $\phi$ are found to be
$$
\phi^i_i=1\, , \quad \phi_{24}^1 = -\frac{1}{3} \, , \quad \phi_{42}^1 = \frac{2}{3} \, , \quad \phi_{44}^3 = \frac{1}{3} \, ,
$$
and from this and~(\ref{noncycAinf}) we can finally obtain the $\Ainf$-products $r_n^{\text{min}}$ in step~(vii). Their coefficients are
\begin{align}\label{cycAinf}
& Q_{11}^1 = -1  \, , \quad && Q_{223}^1 = -2/3  \, , \quad && Q_{2244}^1 = +4/9  \, , \quad && Q_{44444}^1 = -11/27  \nonumber\\
& Q_{12}^2 = -1  \, , \quad && Q_{224}^2 = -2/3  \, , \quad && Q_{2424}^1 = -2/9  \, , \quad &&   \nonumber\\
& Q_{13}^3 = -1  \, , \quad && Q_{232}^1 = +1/3  \, , \quad && Q_{2442}^1 = +2/9  \, , \quad &&   \nonumber\\
& Q_{14}^4 = -1  \, , \quad && Q_{234}^3 = -2/3  \, , \quad && Q_{2444}^3 = +4/9  \, , \quad &&   \nonumber\\
& Q_{21}^2 = -1  \, , \quad && Q_{242}^2 = -1/3  \, , \quad && Q_{3334}^1 = +5/9  \, , \quad &&   \nonumber\\
& Q_{23}^4 = -1  \, , \quad && Q_{243}^3 = -2/3  \, , \quad && Q_{3444}^1 = +5/9  \, , \quad &&   \nonumber\\
& Q_{31}^3 = +1  \, , \quad && Q_{322}^1 = -2/3  \, , \quad && Q_{4224}^1 = +4/9  \, , \quad &&   \nonumber\\
& Q_{32}^4 = +1  \, , \quad && Q_{324}^3 = +1/3  \, , \quad && Q_{4242}^1 = -2/9  \, , \quad &&   \nonumber\\
& Q_{33}^2 = -1  \, , \quad && Q_{334}^1 = -1/3  \, , \quad && Q_{4244}^3 = -2/9  \, , \quad &&   \nonumber\\
& Q_{41}^4 = +1  \, , \quad && Q_{342}^3 = +2/3  \, , \quad && Q_{4344}^1 = +5/9  \, , \quad &&   \nonumber\\
&   \quad && Q_{343}^1 = -1  \, , \quad && Q_{4422}^1 = +4/9  \, , \quad &&   \nonumber\\
&   \quad && Q_{344}^2 = -1/3  \, , \quad && Q_{4424}^3 = +2/9  \, , \quad &&   \nonumber\\
&   \quad && Q_{422}^2 = -2/3  \, , \quad && Q_{4434}^1 = +5/9  \, , \quad &&   \nonumber\\
&   \quad && Q_{423}^3 = -1/3  \, , \quad && Q_{4442}^3 = -4/9  \, , \quad &&   \nonumber\\
&   \quad && Q_{432}^3 = +2/3  \, , \quad && Q_{4444}^2 = +5/9  \, , \quad &&   \nonumber\\
&   \quad && Q_{433}^1 = -1/3  \, , \quad &&    \quad &&   \nonumber\\
&   \quad && Q_{434}^2 = -1  \, , \quad &&    \quad &&   \nonumber\\
&   \quad && Q_{443}^2 = -1/3  \, , \quad &&    \quad &&
\end{align}
and from this we see that $(r_n^{\text{min}})$ is indeed cyclic and unital. We can also check that $(H,r_n^{\text{min}})$ is $\Ainf$-quasi-isomorphic to $(A,r_1,r_2)$ which means that~(\ref{cycAinf}) is the final result and no higher order products are left to be computed. 

Thus we also have determined all the topological string theory amplitudes~(\ref{amplitudes}) in this example as they are always obtained from the coefficients $Q_{a_1\ldots a_n}^a$ by lowering their upper index with the topological metric:
$$
Q_{a_0a_1\ldots a_n} = \sum_a \langle e_{a_0},e_a\rangle_{\text{LG}} \, Q_{a_1\ldots a_n}^a \, .
$$

To obtain the effective superpotential, all that is left to do is to sum up~(\ref{dd}) and~(\ref{cycAinf}) according to
\begin{align*}
\mathcal W_{\text{eff}} & = \sum_{n\geq 2}\frac{1}{n+1} \left\langle e_{a_0}, r_n^{\text{min}}(e_{a_1}\otimes\ldots\otimes e_{a_n})\right\rangle_{\text{LG}} u_{a_0}u_{a_1}\ldots u_{a_n} \\
& = \sum_{n\geq 2}\frac{1}{n+1} (-1)^{\widetilde a_0 + 1}\omega_{a_0a}Q_{a_1\ldots a_n}^a\,u_{a_0}u_{a_1}\ldots u_{a_n} \\
& = \sum_{n\geq 2}\frac{1}{n+1} Q_{a_0 a_1\ldots a_n}\,u_{a_0}u_{a_1}\ldots u_{a_n}
\end{align*}
and we find
$$
\mathcal W_{\text{eff}} = \frac{1}{3}u_3^3 + \frac{5}{6} u_3^2 u_4^2 - \frac{5}{9} u_3 u_4^4 + \frac{11}{162} u_4^6 \, .
$$
After the field redefinition $u_3\mapsto u_3+\frac{2}{3}u_4^2$ and a global rescaling by~$\frac{1}{5}$, this is precisely the same result as in~\cite{hll0402}. 

\vspace{0.2cm}

Our algorithm may in the same way be applied to any other matrix factorisation to obtain amplitudes and superpotentials. Two further simple examples for the latter are
\begin{align*}
\mathcal W_{\text{eff}}\big|_{\left(\begin{tiny}\begin{smallmatrix}0&x^2\\x^2&0\end{smallmatrix}\end{tiny}\right)} & = u_3 u_4^2 - \frac{1}{3} u_3^3 u_4 - \frac{1}{54} u_3^5 \, , \\
\mathcal W_{\text{eff}}\big|_{\left(\begin{tiny}\begin{smallmatrix}0&x^3\\x^3&0\end{smallmatrix}\end{tiny}\right)} & = u_4^2 u_6 + u_4 u_5^2 - u_4 u_5 u_6^2 - \frac{1}{3} u_5^3 u_6 + \frac{1}{6} u_4 u_6^4 - \frac{1}{12} u_5^2 u_6^3 + \frac{5}{108} u_5 u_6^5
\end{align*}
in a suitable choice of basis. 

\vspace{0.2cm}

As a final example we use our method to compute the effective superpotential for the matrix factorisation
$$
\begin{pmatrix}
&&x&0\\
&&0&x^2\\
x^3&0&&\\
0&x^2&&
\end{pmatrix}
$$
which describes the superposition of two branes. If we choose the basis
\begin{align*}
e_1 &=
\begin{pmatrix}
1& & & \\
& 0& & \\
& & 1& \\
& & & 0
\end{pmatrix}, 
\quad
&& e_2 =
\begin{pmatrix}
0& & & \\
& x& & \\
& & 0& \\
& & & x
\end{pmatrix}, \quad
&& e_3 =
\begin{pmatrix}
0& & & \\
& 1& & \\
& & 0& \\
& & & 1
\end{pmatrix}, \\
e_4 &=
\begin{pmatrix}
0&0 & & \\
x& 0& & \\
& & 0&0 \\
& & 1& 0
\end{pmatrix}, 
\quad
&& e_5 =
\begin{pmatrix}
0&1 & & \\
0& x& & \\
& & 0&x \\
& & 0& 0
\end{pmatrix}, \quad
&& e_6 =
\begin{pmatrix}
& & 0&0 \\
& & 0&-x \\
0&0 & & \\
0&x & & 
\end{pmatrix}, \\
e_7 &=
\begin{pmatrix}
& & 0&0 \\
& & 0&-1 \\
0&0 & & \\
0&1 & &
\end{pmatrix}, 
\quad
&& e_8 =
\begin{pmatrix}
& &0 &-1 \\
& & 0&0 \\
0&x & & \\
0&0 & &
\end{pmatrix}, \quad
&& e_9 =
\begin{pmatrix}
& & 0&0 \\
& & -1&0 \\
0&0 & & \\
x&0 & &
\end{pmatrix}, \\
e_{10} &=
\begin{pmatrix}
& & -1&0 \\
& & 0&0 \\
x^2&0 & & \\
0&0 & &
\end{pmatrix}
\end{align*}
and construct a cyclic, unital $\Ainf$-structure starting from $s^6+s^{10}\in\mathcal C^0(B)/\C$, we find 113~non-vanishing coefficients~$Q_{a_1\ldots a_n}^a$ from which we immediately obtain all amplitudes $Q_{a_0\ldots a_n}$. Summing up these amplitudes we arrive at the expression
\begin{align*}
\mathcal W_{\text{eff}} & = u_6 u_7^2 + u_7 u_8 u_9 + u_6 u_8 u_9 u_{10} + u_8u_9u_{10}^2 - \frac{1}{3} u_6^3u_7 - \frac{1}{3}u_6^2u_8u_9 - \frac{1}{45}u_6^5 + \frac{1}{5}u_{10}^5 \\
&\quad\quad-u_4u_5u_{10}^2 + u_4u_5u_6u_{10} - u_2u_5u_8u_{10} + u_2u_4u_8u_{10} + \frac{1}{3}u_2u_4u_6u_8 \\
&\quad\quad-\frac{1}{3}u_2u_5u_6u_9 - \frac{2}{3} u_4u_5u_6^2
\end{align*}
for the effective superpotential in this example.

\subsection*{Conclusion}

In this paper we have shown how to put to use the underlying $\Ainf$-structure of open topological string theory in the case of Landau-Ginzburg models: treated carefully, it allows to compute amplitudes and effective superpotentials algorithmically for any matrix factorisation. The main step was to identify the correct cyclic, unital and minimal $\Ainf$-structure since we saw that a naive construction of minimal models generically produces non-cyclic products for such theories. Reformulating the problem in terms of non-commutative geometry then allowed to treat it much more generally and obtain a theorem whose proof explicitly constructs all cyclic minimal models for any $\Ainf$-algebra. We implemented this general algorithm on a computer and then applied it to matrix factorisations. Apart from the actual computation of amplitudes, this approach also offers an alternative, path-integral-free derivation of the topological metric in open topological Landau-Ginzburg models. 

\subsubsection*{Acknowledgements}

I am very grateful to C.~Lazaroiu for sharing some of his ideas and for very interesting discussions. I also thank M.~Ballard, E.~Segal and especially A.~Recknagel. For their support I thank R.~Streater, the School of Physical Sciences and Engineering of King's College London, and Studienstiftung des deutschen Volkes.

\providecommand{\href}[2]{#2}

\end{document}